\documentstyle[aps,preprint]{revtex}

\def\be{\begin{equation}}
\def\ee{\end{equation}}
\def\bea{\begin{eqnarray}}
\def\eea{\end{eqnarray}}

\begin{document}

\preprint{SOGANG-HEP 268/00}

\date{\today}

\title{Higher dimensional flat embeddings of \\
(2+1) dimensional black holes}

\author{Soon-Tae Hong$^{1}$\footnote{sthong@phya.snu.ac.kr}, 
Yong-Wan Kim$^{1,2}$\footnote{ywkim65@netian.com}, 
and Young-Jai Park $^{1}$\footnote{yjpark@ccs.sogang.ac.kr}}

\address{$^1$Department of Physics and Basic Science Research
Institute, Sogang University, \\C.P.O. Box 1142, Seoul 100-611, Korea \\
$^2$JeungSanDo Research Institute, Sonhwadong 356-8, 
Jung-gu Taejeon, 301-050, Korea}

\maketitle

\begin{abstract}
We obtain the higher dimensional global flat embeddings of
static, rotating, and charged BTZ black holes. 
On the other hand, we also study the similar higher dimensional 
flat embeddings of the (2+1) de Sitter black holes which are the 
counterparts of the anti-de Sitter BTZ black holes.  As a result, 
the charged dS black hole is shown to be embedded in 
(3+2) GEMS, contrast to the charged BTZ one having (3+3) 
GEMS structure.  
\end{abstract}


\section{Introduction}

After Unruh's work \cite{unr}, it has been known that a thermal
Hawking effect on a curved manifold \cite{hawk75} can be looked
at as an Unruh effect in a higher flat dimensional space time.
According to the global embedding Minkowski space (GEMS) approach
\cite{kasner,fro,goenner,ros,n}, several authors 
\cite{des,bec,gon,gib99} recently have shown 
that this approach could yield a
unified derivation of temperature for various curved manifolds such 
as rotating Banados-Teitelboim-Zanelli (BTZ) 
\cite{btz1,btz2,cal,kps,mann93}, Schwarzschild \cite{sch} 
together with its anti-de Sitter (AdS) extension, Reissner-Nordstr\"om 
(RN) \cite{rn,pel} and RN-AdS \cite{kps99}. 

On the other hand, since its pioneering work in 1992, 
the (2+1) dimensional BTZ black hole \cite{btz1,btz2} 
has become one of useful models for realistic black hole physics \cite{cal}.  
Moreover, significant interests in this model have recently increased 
with the novel discovery that the thermodynamics of higher dimensional 
black holes can often be interpreted in terms of the BTZ solution \cite{high}.
It is therefore interesting to study the geometry of (2+1) dimensional 
black holes and their thermodynamics through further investigation.

In this paper we will analyze Hawking and Unruh effects of the
(2+1) dimensional black holes in terms of the GEMS approach. In
section 2 after we briefly recapitulate the known global (2+2)
dimensional embedding of the static and rotating (2+1) BTZ black
holes, we will newly consider the charged static BTZ black hole. 
In section 3, we will also treat the novel global higher 
dimensional flat embeddings of the (2+1) static, rotating, and 
charged de Sitter(dS) black holes, which are the counterpart of
usual BTZ black holes.  
In particular, we will show that the charged dS black hole is 
embedded in (3+2) GEMS, contrast to the charged BTZ one with (3+3) 
GEMS structure.

\section{BTZ Anti-de Sitter Geometry}
\subsection{Static BTZ Space}

In this subsection we begin with a brief recapitulation of the GEMS approach
to temperature given in Ref. \cite{btz2}, 
for the well known (2+1) dimensional uncharged static BTZ black hole 
which is described by the 3-metric
\be
ds^2=N^2dt^2-N^{-2}dr^{2}-r^{2}d{\phi}^{2}
\label{3metric}
\ee
with the lapse function
\be
N^{2}=-M+\frac{r^{2}}{l^{2}}. \ee
Here one notes that this BTZ
space originates from AdS one via the geodesic identification
$\phi=\phi+2\pi$.  The (2+2) AdS GEMS
$ds^{2}=(dz^{0})^2-(dz^{1})^2 -(dz^{2})^2+(dz^{3})^{2}$ is then
given by the coordinate transformations for $r\geq r_{H}$ 
(the extension to $r < r_{H}$ is given in Ref. \cite{btz2}.)
with the event horizon $r_{H}= M^{1/2}l$  as follows
\bea
z^{0}&=&k_H^{-1}\left(\frac{r^{2}-r_{H}^{2}}{l^{2}}\right)^{1/2}\sinh
\frac{r_{H}}{l^2}t, \nonumber \\
z^{1}&=&k_H^{-1}\left(\frac{r^{2}-r_{H}^{2}}{l^{2}}\right)^{1/2}\cosh
\frac{r_{H}}{l^2}t, \nonumber \\
z^{2}&=&l\frac{r}{r_{H}}\sinh\frac{r_{H}}{l}\phi, \nonumber \\
z^{3}&=&l\frac{r}{r_{H}}\cosh \frac{r_{H}}{l}\phi, 
\label{btz0}
\eea
where $k_{H}=r_{H}/l^{2}$ is the Hawking-Bekenstein horizon surface gravity.  
These flat embeddings of the curved spacetime are easily obtained by 
comparing the 3-metric (\ref{3metric}) with $ds^2=\eta_{ab}dz^a dz^b$,
where $a, b=0,\cdots,3$ and $\eta_{ab}=(+,-,-,+)$.

In static detectors ($\phi$, $r=$ const) described by a fixed
point in the ($z^{2}$, $z^{3}$) plane (for example $\phi=0$ gives
$z^{2}=0$, $z^{3}=$const), one can have constant 3-acceleration
\be
a=\frac{r}{l(r^{2}-r_{H}^{2})^{1/2}}
\ee
and constantly accelerated motion in ($z^{0}$,$z^{1}$) 
with the Hawking temperature
\be
T=\frac{a_{4}}{2\pi}=\frac{r_{H}}{2\pi l(r^2-r_{H}^{2})^{1/2}}
\label{ths}
\ee
which yields the relation $a_{4}=(a^{2}-l^{-2})^{1/2}$.  
Here one notes that the above Hawking temperature 
is also given by the relation \cite{hawk75,bek73,wald94}
\be
T=\frac{1}{2\pi}\frac{k_{H}}{g_{00}^{1/2}}. 
\label{g00}
\ee 
Note that in the asymptotic limit the BTZ space approaches to AdS one 
whose acceleration at infinity is given by $a=l^{-1}$ to yield 
zero temperature (no Hawking particle at infinity).   
The Rindler horizon condition $(z^1)^2- (z^0)^2 = 0$ implies $r=r_H$ 
and the remaining embedding constraint yields 
$(z^3)^2 - (z^2)^2 =l^2$ so that the BTZ solution yields a finite Unruh area 
$2\pi r_{H}$ due to the periodic identification of $\phi$ mod $2\pi$ 
\cite{deser99}.  The well-known entropy $2\pi r_{H}$ of the static BTZ space 
is then given by the transverse Rindler area \cite{gib77}.

\subsection{Rotating BTZ Space}
In this subsection we also briefly summarize the results of the GEMS approach 
given in Ref. \cite{btz2,cal,deser99}, for the well known (2+1) dimensional 
uncharged rotating BTZ black hole which is described by the 3-metric
\be
ds^2=N^2dt^2-N^{-2}dr^{2}-r^{2}(d{\phi} + N^{\phi}dt)^{2},
\label{3metricr} 
\ee 
where the lapse and shift functions are
\be
N^{2}=-M+ \frac{r^2}{l^2} + \frac{J^2}{4r^2} ,~~
N^{\phi}=-\frac{J}{2r^2}, 
\ee 
respectively.
Note that for the nonextremal case there exist
two horizons $r_{\pm}(J)$ satisfying the following
equations,
\be
0=-M+ \frac{r_{\pm}^2}{l^2} +
\frac{J^2}{4r_{\pm}^2},
\ee
respectively. Then, without solving these equations explicitly
we can rewrite the mass $M$ and angular momentum $J$ in terms of
these outer and inner horizons as follows
\be
M=\frac{r_{+}^2 + r_{-}^2}{l^2},~~ J=\frac{2r_{+}r_{-}}{l}.
\ee
Furthermore by using these relations, we can rewrite the
lapse and shift functions as
\be
N^2= \frac{(r^2 - r_{+}^2)(r^2 - r_{-}^2)}{r^2 l^2},~~
N^{\phi}=-\frac{r_{+}r_{-}}{r^2 l},
\ee
respectively. Here one notes that this BTZ space originates from
AdS one via the geodesic identification $\phi=\phi+2\pi$.
The (2+2) AdS GEMS $ds^{2}=(dz^{0})^2-(dz^{1})^2 -(dz^{2})^2+(dz^{3})^{2}$ 
is then given by the coordinate transformations for $r\geq r_{+}$ as
follows
\bea
z^{0}&=&k_{H}^{-1}\left(\frac{(r^{2}-r_{+}^{2})(r_{+}^{2}-r_{-}^{2})}
        {r_{+}^{2}l^{2}}\right)^{1/2}\sinh
        \left(\frac{r_{+}}{l^2}t-\frac{r_{-}}{l}\phi\right), \nonumber \\
z^{1}&=&k_{H}^{-1}\left(\frac{(r^{2}-r_{+}^{2})(r_{+}^{2}-r_{-}^{2})}
        {r_{+}^{2}l^{2}}\right)^{1/2}\cosh
        \left(\frac{r_{+}}{l^2}t-\frac{r_{-}}{l}\phi\right), \nonumber \\
z^{2}&=&l\left(\frac{r^{2}-r_{-}^{2}}{r_{+}^{2}-r_{-}^{2}}\right)^{1/2}
       \sinh\left(\frac{r_{+}}{l}\phi-\frac{r_{-}}{l^2}t\right), \nonumber \\
z^{3}&=&l\left(\frac{r^{2}-r_{-}^{2}}{r_{+}^{2}-r_{-}^{2}}\right)^{1/2}
       \cosh\left(\frac{r_{+}}{l}\phi-\frac{r_{-}}{l^2}t\right), 
\label{btz2} 
\eea
where the surface gravity is given as 
$k_{H}=(r_{+}^{2}-r_{-}^{2})/(r_{+}l^{2})$.  
For the trajectories, which follow the Killing vector 
$\xi=\partial_{t}-N^{\phi}\partial_{\phi}$, one can obtain constant 
3-acceleration \cite{deser99}
\be
a=\frac{r^{4}-r_{+}^{2}r_{-}^{2}}{r^{2}l(r^{2}-r_{+}^{2})^{1/2}
  (r^{2}-r_{-}^{2})^{1/2}}.
\ee
and the Hawking temperature \cite{cal,brown}
\be
T=\frac{a_{4}}{2\pi}=\frac{r(r_{+}^{2}-r_{-}^{2})}{2\pi r_{+}l(r^{2}-r_{+}^{2})^{1/2}(r^{2}-r_{-}^{2})^{1/2}}.
\ee
Here these trajectories do not describe pure Rindler motion in the GEMS 
combining accelerated motion in the $(z^{0}, z^{1})$ plane with a spacelike 
motion in $(z^{2}, z^{3})$ \cite{deser99}.  

Finally, the entropy of the rotating BTZ space is given by
$2\pi r_{+}(J)$
which reproduces the uncharged static BTZ black hole entropy $2\pi r_H$ 
in the vanishing $J$ limit.  
Note that all results in this subsection will be useful to analyze 
the dS cases.

\subsection{Charged BTZ Space}

We now newly consider the charged static BTZ black hole solution where
the 3-metric (\ref{3metric}) is described by the charged lapse 
\cite{btz1,zan99}
\be
N^{2}=-M+\frac{r^{2}}{l^{2}}-2Q^{2}\ln r.
\ee
Here the mass $M$ can be rewritten as 
$M=\frac{r_{H}^{2}}{l^{2}}-2Q^{2}\ln r_{H}$ with the horizon
$r_{H}(Q)$, which is the root of $-M+\frac{r^{2}}{l^{2}}-2Q^{2}\ln r=0$.

As in the previous two cases, we can first find the $-r^2d\phi^2$
term in the 3-metric by introducing two coordinates $(z^{3}, z^{4})$ in Eq. 
(\ref{btzch}), giving $-(dz^3)^2+(dz^4)^2=-r^2d\phi^2+\frac{l^2}{r_H^2}dr^2$.
Then, in order to obtain the $N^2dt^2$ term, we make ansatz of two
coordinates, $(z^{0}, z^{1})$ in Eq. (\ref{btzch}), which, together with
the above $(z^3, z^4)$, yields 
\begin{eqnarray} 
\label{temp}
&&(dz^0)^2-(dz^1)^2-(dz^3)^2+(dz^4)^2 \nonumber \\
&&=N^2dt^2-\left(\frac{r_H^2(\frac{r^2}{l^2}-\frac{Q^2r_H}{r})^2}
{(\frac{r_H^2}{l^2}-Q^2)^2(\frac{r^{2}-r_{H}^{2}}{l^{2}}-2Q^{2}
\ln\frac{r}{r_{H}})}-\frac{l^2}{r_H^2}\right)dr^2-r^2d\phi^2.
\end{eqnarray}
Since the combination of $N^{-2}dr^{2}$ and $dr^2$ term in Eq. (\ref{temp}) 
can be separated into positive definite part and negative one as follows 
\begin{eqnarray}
\label{temp2}
&&\left(k_{H}^{-1}Q^{2}
\frac{l[r^{2}+r_{H}^{2}+2r^{2}f(r,r_{H})]^{1/2}}
{r_{H}^{2}r[1-\frac{Q^{2}l^{2}}{r_{H}^{2}}
f(r,r_{H})]^{1/2}}{\rm d}r \right)^2 
-\left(k_{H}^{-1}Q \frac{[2r_{H}^{2}+\frac{r_{H}^{4}+Q^{4}l^{4}}
{r_{H}^{2}}f(r,r_{H})]^{1/2}}
{r_{H}^{2}[1-\frac{Q^{2}l^{2}}{r_{H}^{2}}f(r,r_{H})]^{1/2}}
{\rm d}r\right)^2 \nonumber \\
&& \equiv (dz^2)^2 - (dz^5)^2,
\end{eqnarray}
we can obtain desired flat global embeddings of the corresponding 
curved 3-metric as
\begin{eqnarray}
ds^{2}&=&(dz^{0})^2-(dz^{1})^2
  -(dz^{2})^2-(dz^{3})^{2} +(dz^{4})^{2}+(dz^{5})^{2} \nonumber \\
      &=& N^2dt^2-N^{-2}dr^2-r^2d\phi^2.
\end{eqnarray}
Note that the $z^2$ and $z^5$ are monotonic functions in the range
of $Ql\leq r_H < r$.

As results, we here summarize the (3+3) AdS GEMS given by the
following coordinate transformations with an additional timelike 
dimension $z^{5}$

\bea
z^{0}&=&k_{H}^{-1}\left(\frac{r^{2}-r_{H}^{2}}{l^{2}}-2Q^{2}
\ln\frac{r}{r_{H}}\right)^{1/2}\sinh k_{H}t \nonumber \\
z^{1}&=&k_{H}^{-1}\left(\frac{r^{2}-r_{H}^{2}}{l^{2}}-2Q^{2}
\ln\frac{r}{r_{H}}\right)^{1/2}\cosh k_{H}t \nonumber \\
z^{2}&=&k_{H}^{-1}Q^{2}\int {\rm d}r
\frac{l[r^{2}+r_{H}^{2}+2r^{2}f(r,r_{H})]^{1/2}}
{r_{H}^{2}r[1-\frac{Q^{2}l^{2}}{r_{H}^{2}}
f(r,r_{H})]^{1/2}} \nonumber \\
z^{3}&=&l\frac{r}{r_{H}}\sinh \frac{r_{H}}{l}\phi \nonumber \\
z^{4}&=&l\frac{r}{r_{H}}\cosh \frac{r_{H}}{l}\phi \nonumber \\
z^{5}&=&k_{H}^{-1}Q\int {\rm d}r
\frac{[2r_{H}^{2}+\frac{r_{H}^{4}+Q^{4}l^{4}}{r_{H}^{2}}f(r,r_{H})]^{1/2}}
{r_{H}^{2}[1-\frac{Q^{2}l^{2}}{r_{H}^{2}}f(r,r_{H})]^{1/2}},
\label{btzch}
\eea
where the surface gravity is given by $k_{H}=[(r_{H}/l)^{2}-Q^{2}]/r_{H}$ and
\be
f(r,r_{H})=\frac{2r_{H}^{2}}{r^{2}-r_{H}^{2}}\ln \frac{r}{r_{H}}
\label{frrh}
\ee
which, due to L'Hospital's rule, approaches to unity as $r$ goes to infinity.  

In static detectors ($\phi$, $r=$ const)
described by a fixed point in the ($z^{2}$,$z^{3}$) plane 
(for example $\phi=0$ gives $z^{2}=0$,
$z^{3}=$const), one can have constant 3-acceleration 
\be
a=\frac{r-Q^{2}l^{2}/r}{l[r^{2}-r_{H}^{2}-2Q^{2}l^{2}\ln (r/r_{H})]^{1/2}}
\ee
and constantly accelerated motion in ($z^{0}$,$z^{1}$) 
with the Hawking temperature 
\be
T=\frac{a_{6}}{2\pi}=\frac{r_{H}-Q^{2}l^{2}/r_{H}}{2\pi l[r^2-r_{H}^2-2Q^{2}l^{2}\ln(r/r_{H})]^{1/2}},
\label{thc}
\ee
which is also attainable via the relation (\ref{g00}).  
Note that, in the GEMS where one can have a 
constant Rindler-like accelerated motion, the temperature (\ref{thc}) 
measured by the detector agrees with the temperature given by 
the response function of particle detectors\cite{lif94}.  
Here one can easily check that, in the uncharged limit 
where the spacelike $z^{2}$ and timelike $z^{5}$ dimensions in Eq. 
(\ref{btzch}) vanish, the above coordinate 
transformations are exactly reduced to the previous one (\ref{btz0}) 
for the uncharged static BTZ case.  
Moreover, the desired black hole temperature is given as
\be
T_0=g_{00}^{1/2}T=\frac{(r_{H}/l)^{2}-Q^2}{2 \pi r_H},
\ee
which enters into the black hole thermodynamics relations.
Here one notes that use of incomplete embedding spaces, that cover
only $r>r_H$ (as for example in Ref. \cite{ros}), will lead to observers there
for whom there is no event horizon, no loss of information, and no
temperature.

We now see how the BTZ solution yields a finite Unruh area 
due to the periodic identification of $\phi$ mod $2\pi$.  
The Rindler horizon condition $(z^1)^2- (z^0)^2 = 0$ implies $r=r_H$ 
and the remaining embedding constraints yield
$z^{2}=f_{1}(r)$, $z^{5}=f_{2}(r)$ and $(z^4)^2 - (z^3)^2 =l^2$ 
where $f_{1}(r)$ and $f_{2}(r)$ can be read off from 
Eq. (\ref{btzch}).  The area of the Rindler horizon is now described as  
\[
\int {\rm d}z^{2}{\rm d}z^{3}{\rm d}z^{4}{\rm d}z^{5}\delta(z^{2}-f_{1}(r))\delta(z^{5}-f_{2}(r))
\delta([(z^4)^2-(z^3)^2]^{1/2}-l)
\]
which, after performing trivial integrations over $z^2$ and $z^5$, 
yields the desired entropy of the charged BTZ space
\bea
& &\int^{l\sinh (\pi r_H /l)}_{-l \sinh (\pi r_H /l)}{\rm d}z^{3}
     \int^{[(z^3)^2+l^2]^{1/2}}_{0}{\rm d}z^{4}
     \delta([(z^4)^2-(z^3)^2]^{1/2}-l)\nonumber\\
&=&\int^{l\sinh (\pi r_H /l)}_{-l \sinh (\pi r_H /l)}{\rm d}z^{3}
     \frac{l}{[l^2 +(z^3)^2]^{1/2}}=2\pi r_{H}(Q),
\eea
which reproduces the entropy $2\pi r_{H}$ of the uncharged BTZ case 
in the limit $Q\rightarrow 0$.

It seems appropriate to comment on the minimal extra dimensions 
needed for desired GEMS.  As you may know, spaces of constant curvature can 
be embedded into flat space with only single extra dimension.  This 
is seen in the previous subsections for the static and rotating BTZ 
cases, which are embedded in (2+2)-dimensional spaces.  
On the other hand, since the charged BTZ solution is not locally AdS, 
we have introduced three extra dimensions for desired GEMS.  In the next
section, we will also obtain similar results for the uncharged and 
charged (2+1)-dimensional dS cases. 
\section{ (2+1) de Sitter Black Holes}
\subsection{ Static de Sitter Space}

The static dS black hole \cite{jac84,strom98} is described by the 3-metric 
(\ref{3metric}) with the lapse function
\be
N^{2}=M-\frac{r^{2}}{l^{2}}.
\ee
It arises from dS upon making the geodesic identification
$\phi=\phi+2\pi$. The coordinate transformations to the (3+1) dS
GEMS $ ds^{2}=(dz^{0})^2-(dz^{1})^2 -(dz^{2})^2-(dz^{3})^{2}$ are
for $r\leq r_{H}$ with the event horizon $r_{H}=M^{1/2}l$ 
\bea
z^{0}&=&k_{H}^{-1}\left(\frac{r_{H}^{2}-r^{2}}{l^{2}}\right)^{1/2} 
         \sinh \frac{r_{H}}{l^2}t, \nonumber\\
z^{1}&=&k_{H}^{-1}\left(\frac{r_{H}^{2}-r^{2}}{l^{2}}\right)^{1/2} 
         \cosh \frac{r_{H}}{l^2}t, \nonumber\\
z^{2}&=&l\frac{r}{r_{H}}\sin \frac{r_{H}}{l}\phi, \nonumber\\
z^{3}&=&l\frac{r}{r_{H}}\cos \frac{r_{H}}{l}\phi,
\label{zzz0}
\eea
where the constant
$r_{H}$ are related to the mass.  

Even though there is no longer a one to one mapping 
between the dS GEMS and the BTZ like dS space due to the $\phi$ 
identification, following a detector motion with certain initial condition 
such as $\phi(t=0)=0$ still yields a unique trajectory in the embedding space.
If the detector trajectory maps into an Unruh one in the dS GEMS 
without ambiguity, then one can exploit it to evaluate temperature.

Now let us consider static detectors ($\phi$, $r$ $=$ const). 
These detectors have constant 3-acceleration 
\be
a=\frac{r}{l(r_{H}^2-r^{2})^{1/2}},
\label{as}
\ee
and are described by a fixed point in the ($z^{2}, z^{3}$) plane 
(for example $\phi=0$ gives $z^{2}=0,$ $z^{3}=$const), 
to yield constantly accelerated motion in ($z^{0},z^{1}$) 
with the Hawking temperature
\be
T=\frac{a_{4}}{2\pi}=\frac{r_{H}}{2\pi l(r_{H}^2-r^2)^{1/2}},
\label{a4s}
\ee 
which is connected with $a$ by the relation $a_{4}=(a^{2}+l^{-2})^{1/2}$.  
Thus, in the GEMS described in (\ref{zzz0}), we have a 
constant Rindler-like accelerated motion and the temperature (\ref{a4s}) 
measured by the detector, which agrees with the temperature given by 
the response function of particle detectors \cite{lif94}.  
We also obtain the entropy $2\pi r_{H}$ of the static dS space as 
in the BTZ case.

\subsection{Rotating de Sitter Space}

The rotating Kerr-dS black hole\cite{gib77,mip} is 
described by the 3-metric (\ref{3metricr}) with the lapse and shift functions
\be
N^{2}=M- \frac{r^2}{l^2} + \frac{J^2}{4r^2} ,~~
N^{\phi}=-\frac{J}{2r^2}. 
\ee
Similarly to the rotating BTZ black hole case, we can also rewrite 
the mass $M$ and angular momentum $J$ in terms of outer and inner horizons 
$r_{\pm}(J)$ as follows
\be
M=\frac{r_{+}^2 - r_{-}^2}{l^2},~~ J=\frac{2r_{+}r_{-}}{l}.
\label{mjds}
\ee
Furthermore, by using these relations, we can obtain the lapse and shift 
functions 
\be
N^2= \frac{(r_{+}^2 - r^2)(r^2 +r_{-}^2)}{r^2 l^2},~~
N^{\phi}=-\frac{r_{+}r_{-}}{r^2 l},
\ee
respectively.  It arises from dS upon making the geodesic identification
$\phi=\phi+2\pi$. The coordinate transformations to the (3+1) dS
GEMS $ ds^{2}=(dz^{0})^2-(dz^{1})^2 -(dz^{2})^2-(dz^{3})^{2}$ are
for $r\leq r_{+}$ 
\bea
z^{0}&=&k_{H}^{-1}\left(\frac{(r_{+}^{2}+r_{-}^{2})(r_{+}^{2}-r^{2})}
        {r_{+}^{2}l^{2}}\right)^{1/2} \sinh \left(\frac{r_{+}}{l^2}t-
         \frac{r_{-}}{l}\phi \right), \nonumber\\
z^{1}&=&k_{H}^{-1}\left(\frac{(r_{+}^{2}+r_{-}^{2})(r_{+}^{2}-r^{2})}
        {r_{+}^{2}l^{2}}\right)^{1/2} \cosh \left(\frac{r_{+}}{l^2}t-
        \frac{r_{-}}{l}\phi \right), \nonumber\\
z^{2}&=&l\left(\frac{r^{2}+r_{-}^{2}}{r_{+}^{2}+r_{-}^{2}}\right)^{1/2}
        \sin \left( \frac{r_{+}}{l}\phi +\frac{r_{-}}{l^2}t\right), \nonumber\\
z^{3}&=&l\left(\frac{r^{2}+r_{-}^{2}}{r_{+}^{2}+r_{-}^{2}}\right)^{1/2}
        \cos \left( \frac{r_{+}}{l}\phi +\frac{r_{-}}{l^2}t\right),
\label{zzz}
\eea
where the constants $r_{\pm}(J)$ are related to the mass and 
angular momentum as in Eq. (\ref{mjds}). 

Similar to the rotating BTZ case, for the trajectories which follow the 
Killing vector $\xi=\partial_{t}-N^{\phi}\partial_{\phi}$, we obtain constant 
3-acceleration 
\be
a=\frac{r^{4}+r_{+}^{2}r_{-}^{2}}{r^{2}l(r_{+}^{2}-r^{2})^{1/2}
   (r^{2}+r_{-}^{2})^{1/2}},
\ee
and the Hawking temperature
\be
T=\frac{a_{4}}{2\pi}=\frac{r(r_{+}^{2}+r_{-}^{2})}{2\pi r_{+}l
    (r_{+}^{2}-r^{2})^{1/2}(r^{2}+r_{-}^{2})^{1/2}}.
\ee
Note that as in the rotating BTZ black hole these trajectories do not 
describe pure Rindler motion in the GEMS.  
On the other hand, the entropy $2\pi r_{+}(J)$ 
of the rotating Kerr-dS space also reproduces the uncharged static dS black 
hole entropy $2\pi r_{H}$ in the vanishing $J$ limit.  

\subsection{Charged de Sitter Space}

We now consider the charged static dS black hole solution where
the 3-metric (\ref{3metric}) is described by the charged lapse
\be
N^{2}=M-\frac{r^{2}}{l^{2}}-2Q^{2}\ln r.
\ee
Here the mass $M$ can be rewritten as $M=\frac{r_{H}^{2}}{l^{2}}+
2Q^{2}\ln r_{H}$ with the horizon
$r_{H}(Q)$, which is the root of $M-\frac{r^{2}}{l^{2}}-2Q^{2}\ln r=0$.

After similar algebraic manipulation by following the previous steps 
described in Sec. II.C, we obtain the (3+2) dS GEMS 
$ds^{2}=(dz^{0})^2-(dz^{1})^2-(dz^{2})^2-(dz^{3})^{2} +(dz^{4})^{2}$ 
given by the coordinate transformations with only one additional 
timelike dimension $z^{4}$, in contrast to the BTZ case where one
needs to require additionally one spacelike and one timelike dimensions
\bea
z^{0}&=&k_{H}^{-1}\left(\frac{r_{H}^{2}-r^{2}}{l^{2}}-2Q^{2}
\ln\frac{r}{r_{H}}\right)^{1/2}\sinh k_{H}t, \nonumber \\
z^{1}&=&k_{H}^{-1}\left(\frac{r_{H}^{2}-r^{2}}{l^{2}}-2Q^{2}
\ln\frac{r}{r_{H}}\right)^{1/2}\cosh k_{H}t, \nonumber \\
z^{2}&=&l\frac{r}{r_{H}}\sin \frac{r_{H}}{l}\phi, \nonumber \\
z^{3}&=&l\frac{r}{r_{H}}\cos \frac{r_{H}}{l}\phi, \nonumber \\
z^{4}&=&k_{H}^{-1}Q\int {\rm d}r
\frac{\{Q^{2}l^{2}[r^{2}+r_{H}^{2}+2r^{2}f(r,r_{H})]
+r^{2}[2r_{H}^{2}+\frac{r_{H}^{4}+Q^{4}l^{4}}{r_{H}^{2}}f(r,r_{H})]\}^{1/2}}
{r_{H}^{2}r(1+\frac{Q^{2}l^{2}}{r_{H}^{2}}f(r,r_{H}))^{1/2}},\nonumber \\
\label{dsch}
\eea
where the surface gravity is given by $k_{H}=[(r_{H}/l)^{2}+Q^{2}]/r_{H}$ 
and $f(r,r_{H})$ is given by Eq. (\ref{frrh}).  
Here one can easily check that, in the uncharged limit, the above coordinate 
transformations are reduced to the previous one (\ref{zzz0}) 
for the uncharged static dS case.  Note that in dS space we need only one 
additional dimension $z^{4}$ since the $Q^{2}$ term in the 
numerator has the same sign with respect to the second term, differently from 
the charged static BTZ case where we have the opposite relative sign between 
these two terms to yield two additional dimensions, namely the spacelike 
$z^{2}$ and timelike $z^{5}$ in (\ref{btzch}). 

In static detectors ($\phi$, $r=$ const)
described by a fixed point in the ($z^{2}$, $z^{3}$) plane 
(for example $\phi=0$ gives $z^{2}=0$,
$z^{3}=$const), one can have constant 3-acceleration 
\be
a=\frac{r+Q^{2}l^{2}/r}{l[r_{H}^{2}-r^{2}-2Q^{2}l^{2}\ln (r/r_{H})]^{1/2}}
\ee
and the Hawking temperature in constantly accelerated motion in 
($z^{0}$,$z^{1}$)  
\be
T=\frac{a_{5}}{2\pi}=\frac{r_{H}+Q^{2}l^{2}/r_{H}}{2\pi l
   [r_{H}-r^2-2Q^{2}l^{2}\ln(r/r_{H})]^{1/2}}.
\ee
In the GEMS one can thus have a constant Rindler-like accelerated motion 
and the above Hawking temperature measured by the detector.  
Note that, in the uncharged static limit $Q\rightarrow 0$, the above 
3-acceleration and Hawking temperature are reduced to the previous ones 
(\ref{as}) and (\ref{a4s}).  The desired black hole temperature is then 
given as
\be
T_0=g_{00}^{1/2}T=\frac{(r_{H}/l)^{2}+Q^2}{2 \pi r_H}.
\ee
Note that the entropy $2\pi r_{H}(Q)$ of the charged dS black hole is also 
given by the Rindler horizon condition. This is also reduced to
the entropy $2\pi r_{H}$ for the uncharged static dS case in the limit of 
$Q \rightarrow 0$. 

\section{Conclusion}

In conclusion, we have shown that Hawking thermal properties map
into their Unruh equivalents by globally embedding various curved (2+1) 
dimensional BTZ and dS spaces into higher dimensional flat ones.  
The relevant curved space detectors become Rindler ones, 
whose temperature and entropy reproduce the originals.  
It would be interesting to consider other interesting
applications of GEMS, for example to superradiance in rotating
Kerr type geometries \cite{btz1,kps,ker,hkp} or Chan's new classes 
of static BTZ black hole solution due to a
chosen asymptotically constant dilation and scalar \cite{chan}.

Finally, it seems appropriate to comment on the rotating version of 
charged BTZ black hole.  As pointed out by several authors, 
if one includes electric charge $Q$,
the solution of the field equations is attainable only when the angular 
momentum $J$ vanishes \cite{cal,kama}. 
However, very recently several authors \cite{mar99} have analyzed the case 
when all three 'hairs' $M,J,$ and $Q$ are different from zero 
although they have treated in some restricted ranges.  
Therefore, it is very interesting to show whether the solutions of 
the rotating charged BTZ and dS black holes may be obtained or 
not in terms of GEMS approach through further investigation.

\acknowledgements{
We would like to thank W.T. Kim and M.I. Park for helpful discussions.  
We acknowledge financial support from the Ministry of 
Education, BK21 Project No. D-0055, 1999.}


\end{document}